\documentclass[journal]{IEEEtran}
\pdfoutput=1

\ifCLASSINFOpdf
  \usepackage[pdftex]{graphicx}
  \graphicspath{{figs/}}
  \DeclareGraphicsExtensions{.pdf,.jpeg,.png}
\fi

\usepackage{microtype}
\usepackage{amsmath}

\usepackage{url}
\usepackage{booktabs}
\usepackage{multirow}
\usepackage{color}

\usepackage{pifont}
\newcommand{\cmark}{\ding{51}}
\newcommand{\xmark}{\ding{55}}
\newcommand{\smark}{\ding{86}}

\newcommand{\code}[1]{\texttt{\small #1}}

\usepackage[dvipsnames]{xcolor}
\usepackage{listings}
\usepackage{courier}

\lstdefinelanguage{SWIG}{%
  morekeywords={typemap, module, include, apply, template, inline, noblock, SWIGTYPE},
  morecomment=[l]{//},
}

\usepackage[colorinlistoftodos,prependcaption,textsize=small]{todonotes}
\usepackage{xargs}
\usepackage{soul}
\newcommandx{\fix}[2]{\todo[linecolor=RedOrange,backgroundcolor=RedOrange!25,bordercolor=RedOrange,inline]{\textbf{#1: }#2}}

\newcommandx{\changed}[1]{#1}

\begin{document}

\title{Automated Fortran--C++ Bindings \\ for Large-Scale Scientific Applications}

\author{Seth~R.~Johnson,
        Andrey~Prokopenko,
        and~Katherine~J.~Evans
\thanks{This manuscript has been authored by UT--Battelle, LLC, under Contract No.
DE-AC05-00OR22725 with the U.S. Department of Energy. The United States
Government retains and the publisher, by accepting the article for publication,
acknowledges that the United States Government retains a non-exclusive, paid-up,
irrevocable, world-wide license to publish or reproduce the published form of
this manuscript, or allow others to do so, for United States Government
purposes.}
}

\markboth{arXiv pre-print}%
{Bare Demo of IEEEtran.cls for IEEE Journals}

\maketitle


\begin{abstract}
Although many active scientific codes use modern Fortran, most contemporary
scientific software \emph{libraries} are implemented in C and C++. Providing their numerical, algorithmic, or data management features to Fortran codes requires writing and maintaining substantial amounts of glue code.

This article introduces a tool that automatically generates native Fortran 2003
interfaces to C and C++ libraries. The tool supports C++ features that have no
direct Fortran analog, such as templated functions and exceptions. A set of
simple examples demonstrate the utility and scope of the tool, and timing
measurements with a mock numerical library illustrate the minimal performance
impact of the generated wrapper code.

\end{abstract}




%
\IEEEpeerreviewmaketitle

\ifCLASSOPTIONcaptionsoff
  \newpage
\fi

\section{Introduction}\label{s:intro}

\IEEEPARstart{F}{ortran} has a long history in scientific programming and is
still in common use today~\cite{decyk2007fortran} in application codes for
climate science \cite{e3sm}, weather forecasting \cite{um}, chemical
looping reactors \cite{mfix}, plasma physics, and other fields.
As a domain-specific language for scientific computing, Fortran enables
modernization and improvement of application codes by the publication of new
standards specifications that define new features and intrinsic functions.
Unfortunately, these specifications take years to become available to Fortran
users: as of 2018, only six of eleven surveyed compilers fully implement even
half of Fortran 2008's new features \cite{chivers}.
Another approach to language extensibility is the distribution of libraries,
which can be developed and deployed much more rapidly than compilers. However,
most contemporary compiled-language scientific and computing libraries
are written in C and C++, and
their capabilities are either unavailable to Fortran users or exposed through
fragile C interface code.
Providing Fortran application developers with robust bindings to
high-performance C++ libraries will substantially and rapidly enrich their
toolset.

As available computing power increases, more scientific codes are improving
their simulations' fidelity by incorporating additional physics. Often
multiphysics codes rely on disparate pieces of scientific software written in
multiple programming languages. Increasingly complex simulations require
improved multi-language interoperability.
Furthermore, the drive to exascale scientific computing motivates the
replacement of custom numerical solvers in Fortran application code with
modern solvers, particularly those with support for distributed-memory
parallelism and device/host architectures.

The increasing necessity for robust, low-maintenance Fortran bindings\changed{---for
multi-language application developers and scientific library authors---}demands a
new tool for coupling Fortran applications to existing C and C++ code. This
article introduces such a tool, implemented as a new extension to the Simplified
Wrapper and Interface Generator (SWIG) tool~\cite{swig2003}. A set of simple
examples demonstrate the utility and scope of SWIG-Fortran, and timing measurements with a mock numerical library illustrate the minimal performance
impact of the generated wrapper code.

\section{Background}

For decades, Fortran application developers and domain scientists have required
capabilities that can only be implemented in a systems programming language such
as C. The MPI specification, which declares both C and Fortran interfaces,
provides a clear example: the MPICH implementation was written in C and to this
day uses a custom tool to automate the generation of a Fortran interface and the
underlying C binding routines \cite{gropp1996}.

Over the years, many attempts have been made to build a generic tool to generate
Fortran bindings to existing C C++ code.
Early efforts explored manual generation of encapsulated \emph{proxy}
Fortran interfaces to C++ classes using Fortran 95 \cite{Gra1999}.
Some C scientific libraries such as Silo, HDF5, and
PETSc include custom tools to generate Fortran from their own
header files. Most of these tools use non-portable means to
bind the languages with the help of configuration scripts, because they were
initially developed before widespread
support for the Fortran 2003 standard~\cite{f2003}, which added features for
standardized interoperability with ISO C code.

Some newer software projects, such as the first iteration of a Fortran
interface~\cite{morris2012} to the Trilinos~\cite{trilinos} numerical library
that motivated this work, use Fortran 2003 features but are limited to manually
generated C interfaces to C++ code, with hand-written Fortran shadow
interfaces layered on those interfaces. The Babel tool~\cite{Epp2012} can
\emph{automatically} generate data
structures and glue code for numerous languages (including C++ and Fortran 2003)
from a custom interface description language, but it is suited for data
interoperability in a polyglot application rather than for exposing existing C
and C++ interfaces to Fortran.

Any practical code generation tool must be able to parse and interact with
advanced language features, such as C++ templates which are critical to today's
parallel scientific libraries.  The advent of GPU accelerators further
complicates inter-language translation by the data sharing imposed by its device/host
dichotomy. The maturity and flexibility of SWIG allows us to address
these and other emergent concerns in a generic tool for generating high-performance,
modern Fortran interfaces from existing C and C++ library header files with
minimal effort from the C++ library developers and no effort on the downstream
Fortran application developers.


\section{Methodology}\label{s:methodology}

The core functionality of SWIG is to parse library header files and generate
C-linkage wrapper interfaces to each function. It composes
these interfaces out of small code snippets called ``typemaps,'' responsible for
converting a type in C++ to a different representation of that type in a target
language.  The new Fortran target language comprises a library of these
typemaps and a C++ ``language module'' compiled into the SWIG executable.

\begin{figure}[htb]
  \centering
  \includegraphics{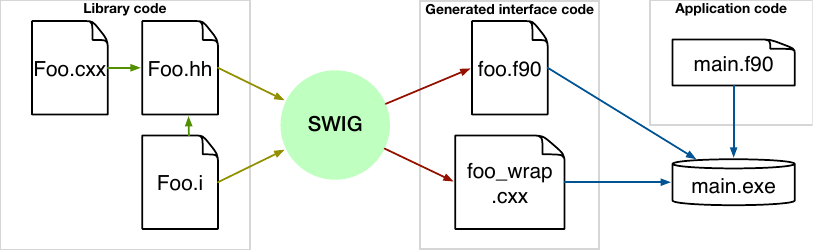}
  \caption{Dependency flow for SWIG-generated code. Green arrows symbolize
    ``includes'', yellow arrows are ``read by,'' red arrows are ``generates,''
  and blue arrows are ``compiled and linked into.''}\label{f:swig_data}
\end{figure}

SWIG generates wrappers only for code specified in an interface file, which can
direct SWIG to process specified C and C++ header files. The interface file
also informs SWIG of additional type conversions and code transformations to be
made to the modules, procedures, or classes to be wrapped. Each invocation of
SWIG-Fortran with an interface file creates two source files: a C++ file with
C-linkage wrapper code that calls the C++ libraries, and a Fortran module file
that calls the C-linkage wrapper code (Fig.~\ref{f:swig_data}).
These generated source files must be compiled and linked together against the
C++ library being wrapped, and the resulting Fortran module can be directly used
by downstream user code. The generated files can be included in a C++ library
distribution without requiring library users or application developers to
install or even have any knowledge of SWIG.

As SWIG processes an interface file (or a header file referenced by that file),
it encounters C and C++ declarations that may generate several pieces of code.
Functions, classes, enumerations, and other declarations generate user-facing
Fortran equivalents and the private wrappers that transform data to pass it
between the Fortran user and the C++ library. These type-mapping transformations
are enabled by the Fortran 2003 standard, which mandates a set of interoperable
data types between ISO C and Fortran, defined in the \code{ISO\_C\_BINDING}
intrinsic module.

\subsection{Type conversions}\label{s:type_conversions}

\changed{%
SWIG-Fortran can pass all ISO C compatible data types between C++ and
Fortran without copying or transforming the data. Additional typemaps included
with the SWIG library provide transformations between C++ types and Fortran
types that are analogous but not directly compatible.}
These advanced type transformation routines shield Fortran application users
from the complexities of inter-language translation.

Consider the character string, which for decades has complicated C/Fortran binding
due to its different representation by the two languages.
The size of a C string is determined by counting the number of characters until
a null terminator \code{\textbackslash0} is encountered, but Fortran string
sizes are fixed at
allocation. The Fortran ISO C binding rules prohibit Fortran strings from being
directly passed to C; instead, SWIG-Fortran injects code to convert a string to a
zero-terminated array of characters, which \emph{can} interact with C.
\changed{A small C-bound struct containing the array's length and the address of its
initial element is passed to the C++ wrapper code, which then can instantiate a
\code{std::string} or pass a \code{char*} pointer to the C/C++ library.}
Returning a string from C++ similarly passes a pointer and length through the C
wrapper layer. The Fortran wrapper code creates an \code{allocatable} character
array of the correct size, copies in the data from the character array, and frees
the C pointer if necessary. Thus, C/C++ library routines that accept or return
strings can be called using \emph{native} types familiar to the Fortran user
without any knowledge of null terminators.

Another notable scalar type conversion defined by SWIG-Fortran is boolean
value translation. In C and C++, the \code{bool} type is defined to be
\emph{true} if
nonzero and \emph{false} if zero, whereas Fortran logical values are
defined to be true if the least significant bit is 1 and false otherwise.
The automated wrapper generation frees developers from having to understand
that the value 2 may be \code{true} in C but \code{false} in Fortran.

Finally, the SWIG typemap system also allows multiple C++ function arguments to
be converted to a single argument in the target language. This
allows a multi-argument C++ function
\begin{lstlisting}[language=C++]
double cpp_sum(const double* arr, std::size_t len);
\end{lstlisting}
to generate a Fortran function that accepts a native Fortran array:
\begin{lstlisting}[language=Fortran]
function cpp_sum(data) result(swig_result)
  real(C_DOUBLE) :: swig_result
  real(C_DOUBLE), dimension(:), intent(IN), target :: data
end function
\end{lstlisting}
by creating temporary arguments using the intrinsic Fortran \code{SIZE} and
\code{C\_LOC} functions.

Advanced typemaps can be constructed to perform other transformations on the
input to facilitate the translation of C++ APIs into forms familiar to Fortran
application developers. For example, the wrappers may increment input parameters
by 1 so that library users can continue using the idiomatic 1-offset indexing of
Fortran rather than counting from 0 as in C++.

\subsection{Functions}

Functions in C/C++ are \emph{procedures} in Fortran.  A function in C/C++ with a
\code{void} return value will translate to a \code{subroutine} in Fortran, and a function
returning anything else will yield a Fortran \code{function}.

Each function processed by SWIG-Fortran generates a single C-linkage shim function in
the C++ file. This thin wrapper converts the
function's arguments from Fortran- and C-compatible datatypes to the function's
actual C++ argument types, calls the function, and converts the result back to a
Fortran-compatible datatype. The wrapper function also implements other optional
features such as exception handling and input validation.

\begin{figure}[htb]
  \centering
  \includegraphics[width=1.75in]{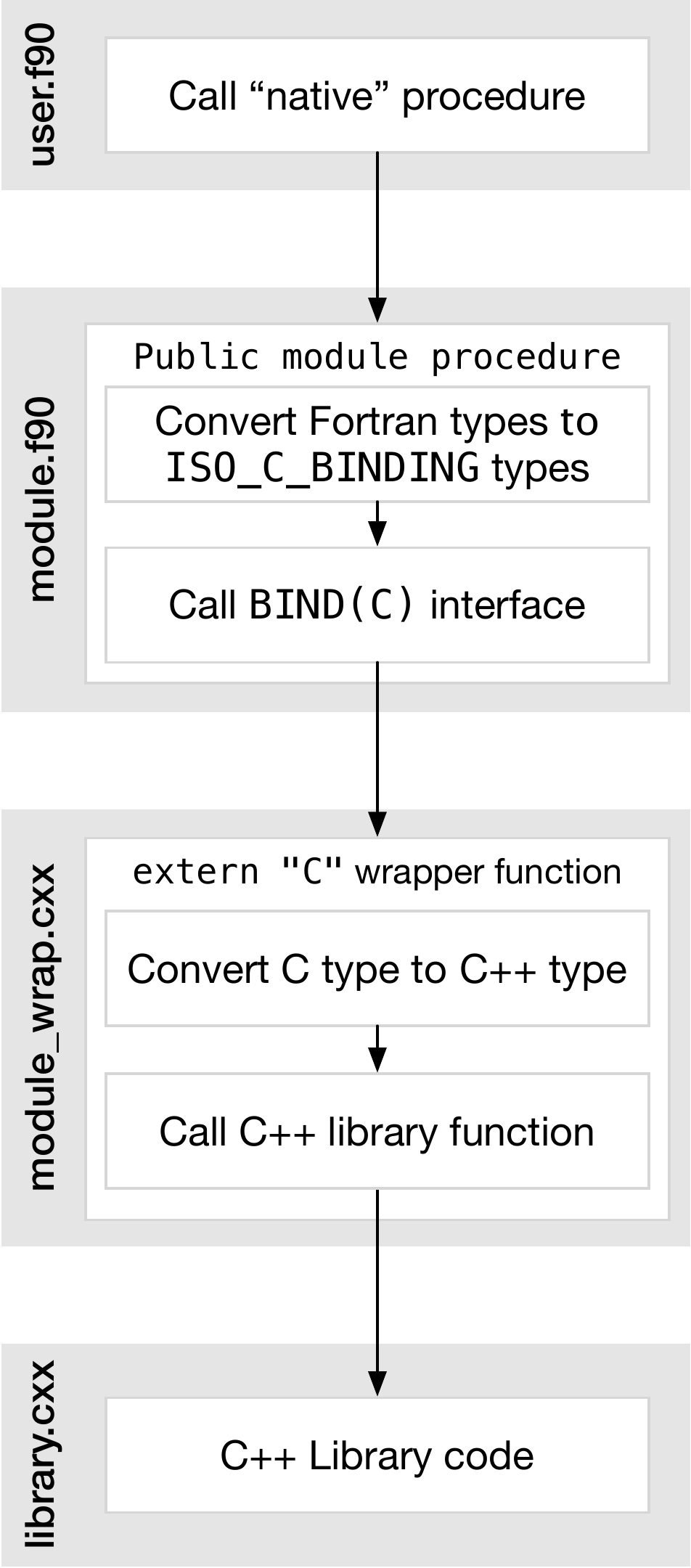}
  \caption{Program flow for calling a C++ library from Fortran through
  SWIG-generated wrappers.}\label{f:code_flow}
\end{figure}

In the corresponding \code{.f90} module file, SWIG-Fortran generates a private,
\code{bind(C)}-qualified declaration of the C wrapper in an \code{INTERFACE}
block.  This interface is called by a public Fortran shim procedure that translates
native Fortran datatypes to and from the C interface datatypes.
Figure~\ref{f:code_flow} demonstrates the control flow for a Fortran user
calling a C++ library function through the SWIG-generated module and wrapper
code.

If a function is overloaded, SWIG-Fortran generates a unique, private Fortran shim procedure
for each overload. These procedures are then combined under a
\emph{separate module procedure} that is given a public interface with the original
function's name. For example, an overloaded free function \code{myfunc} in C++ will
generate two private procedures and add an interface to the module
specification:
\begin{lstlisting}[language=Fortran]
public :: myfunc
interface myfunc
  module procedure myfunc__SWIG_0, myfunc__SWIG_1
end interface
\end{lstlisting}
\changed{%
Since Fortran does not allow a module procedure or generic interface to contain
both functions (which return an object) \emph{and} subroutines (which return
nothing), SWIG-Fortran detects, ignores, and warns about such incompatible
overloaded C++ functions, e.g.,}
\begin{lstlisting}[language=c++]
void overloaded();
int overloaded(int);
\end{lstlisting}

Templated functions are supported, but since the wrapper code is generated
without any knowledge of the downstream Fortran user's code, each template
instantiation must be explicitly specified in the SWIG interface file:
\begin{lstlisting}[language=SWIG]
// C++ library function declaration:
template<class T> void do_it(T value);
// SWIG template instantiations:
%template(do_it_int) do_it<int>;
%template(do_it_real) do_it<float>;
\end{lstlisting}
The two instantiated subroutines can be called from user Fortran code on
integers and single-precision floating-point values, but since the example does
not instantiate on \code{bool}, the user's Fortran code cannot call
\code{do\_it} with a logical argument. Note that the chosen names for
template instantiations (\code{do\_it\_int} and \code{do\_it\_real}) have no
bearing on the way they are being called from a Fortran code, which is through
the original \code{do\_it} name.

The restriction of SWIG-time instantiation may be particularly limiting for
library functions that accept generic functors%
\changed{%
, including the C++11 lambda functions that are crucial to programming models such as Kokkos and RAJA.
However, SWIG's support for function pointer conversion does enable some
flexibility for user codes.} Instantiating a function template using a
\emph{function pointer} as the predicate operator allows the function to be used
with arbitrary functions with a fixed type signature:
\begin{lstlisting}[language=SWIG]
%{
template<class T, class BinaryPredicate>
bool compare(T lhs, T rhs, BinaryPredicate cmp) {
  return cmp(lhs, rhs);
}
%}
typedef bool (*cmp_int_funcptr)(int, int);
%template(compare) compare<int, cmp_int_funcptr>;
\end{lstlisting}
This instantiated \code{compare} function can be used in Fortran with an
arbitrary comparator that accepts integer arguments:
\begin{lstlisting}[language=Fortran]
result = compare(123_c_int, 100_c_int, c_funloc(my_comparator))
\end{lstlisting}
if the user has defined a function such as
\begin{lstlisting}[language=Fortran]
function my_comparator(left, right) bind(C) &
    result(is_less)
  use, intrinsic :: ISO_C_BINDING
  integer(C_INT), intent(in), value :: left
  integer(C_INT), intent(in), value :: right
  logical(C_BOOL) :: is_less

  is_less = modulo(left, 100_c_int) &
            < modulo(right, 100_c_int)
end function
\end{lstlisting}

This limited capability for inversion of control (i.e., C++ library code calling
Fortran application code) will be extended and documented in future work.

\subsection{Automated class generation}

Like the thin wrappers generated for procedures, SWIG also creates thin
proxy wrappers for C++ classes. Here, each Fortran \emph{derived type} mirrors a C++ class,
contains \emph{type-bound procedures} that mirror C++ member functions, and stores a
pointer to the C++ class with memory management flags. Its assignment and memory
management semantics mimic that of an allocatable Fortran pointer. SWIG supports single
inheritance by generating a derived type with the \code{EXTENDS} attribute, so
functions accepting a \code{class(Base)} argument will also accept a
\code{class(Derived)}.

Proxy types are instantiated via a module interface function that shares the
name of the derived type, giving construction the same syntax as in other
high-level languages and in other modern Fortran idioms \cite{rouson2012}.
Instances are nullified (and deallocated if owned) by calling the \code{release}
procedure.

The same proxy type can be used as an interface to a C++ class
pointer, const reference, or value; and it must be able to correctly transfer
ownership during assignment or when being returned from a wrapped C++ function.
To support this variety of similar but distinct use cases in a single instance
of a Fortran type, the proxy type stores a bit field of flags alongside the pointer to
the C++ object. One bit denotes ownership, another marks the instance as a C++
\code{rvalue}, and the third bit is set if the instance is \code{const}. The
\code{rvalue} bit is set only when returning a value from a function. A custom
Fortran \code{assignment(=)} generic function \emph{transfers} ownership when
this bit is set.

The following block of code demonstrates the assignment semantics; it neither
leaks nor double-deletes memory.
\begin{lstlisting}[language=Fortran,numbers=right,numbersep=-10pt,stepnumber=1,numberstyle=\tiny,firstnumber=1]
  type(Foo) :: owner, alias
  owner = Foo(2)
  owner = Foo(3)
  alias = owner
  call alias%release()
  call owner%release()
\end{lstlisting}
In line 2, a new temporary \code{Foo} object is created and returned with the
\code{rvalue} and \code{own} flags. The assignment operation in line 2 transfers
ownership from that temporary object and assigns it to the \code{owner}
instance, replacing the initial value of a null C pointer, and clearing the
\code{rvalue} flag. Line 3 creates another object, but when the SWIG-generated
assignment operator is called, it first \emph{deletes} the original \code{Foo}
object before capturing the new one. Without the special assignment operator,
memory would be leaked. The next assignment (line 4) copies the underlying C
pointer but not the memory flag, so that \code{alias} is a non-owning pointer to
the same object as \code{owner}. Line 5 clears the pointer but does not call any
destructor because it did not own the memory. The final line actually destroys
the underlying object because its ownership flag is set.

This methodology is an alternative to implementing the Fortran proxy type as a
shared pointer that relies on the Fortran \code{FINAL} feature
\cite{rouson2012}. This feature is intentionally avoided because it is not
implemented (or buggy) in some recent compilers, sixteen years after the
specification of Fortran 2003 \cite{chivers}.

\subsection{Exception handling}

Since Fortran has no exception handling, any uncaught C++ exception from a
wrapped library call will immediately terminate the program.
With SWIG's \code{\%exception} feature, C++ exceptions
can be caught and handled by the Fortran code by setting and clearing an
integer flag. For example, assuming that one wants to use a conservative
square root function \code{careful\_sqrt} that throws an exception when a given
number is non-positive, the Fortran code could look like this:
\begin{lstlisting}[language=Fortran]
use except, only : careful_sqrt, ierr, get_serr
call careful_sqrt(-4.0)
if (ierr /= 0) then
  write(0,*) "Got error ", ierr, ": ", get_serr()
  ierr = 0
endif
\end{lstlisting}
where \code{ierr} is a nonzero error code, and \code{get\_serr()} returns a string
containing the exception message. This approach allows Fortran code to
gracefully recover from exception-safe C++ code. For example, if a C++ numeric
solver throws an error if it fails to converge, the Fortran application would
be able to detect the failure, print a message, and write the unconverged
solution.

\subsection{HPC-oriented features}

The SWIG Fortran target language \changed{is able to wrap CUDA}
kernels using the Thrust C++ interface and use the resulting code with Fortran
OpenACC kernels. The implementation is designed to avoid the performance
penalty of copying between the host and device inside the wrapper layer: the
underlying device data pointer is seamlessly handed off between C++/CUDA and
Fortran.

Here is an example SWIG module that wraps the \code{thrust::sort} function to enable
sorting on-device data using a highly optimized kernel:
\begin{lstlisting}[language=SWIG]
%module thrustacc

%include <openacc.i>
%include <thrust.i>
%{
#include <thrust/sort.h>
%}

%inline %{
template<typename T>
static void thrust_sort(thrust::device_ptr<T> DATA, size_t SIZE) {
  thrust::sort(DATA, DATA + SIZE);
}
%}

%template(sort) thrust_sort<float>;
\end{lstlisting}
The corresponding test code simply calls the SWIG-generated \code{sort} function:
\begin{lstlisting}[language=Fortran]
program test_thrustacc
  use thrustacc, only : sort
  implicit none
  integer, parameter :: n = 64
  integer :: i
  integer :: failures = 0
  real, dimension(:), allocatable :: a
  real :: mean

  ! Generate N uniform numbers on [0,1)
  allocate(a(n))
  call random_number(a)
  write(0,*) a

  !$acc data copy(a)
    !$acc kernels
      do i = 1,n
        a(i) = a(i) * 10
      end do
    !$acc end kernels
    call sort(a)
  !$acc end data

  write(0,*) a
end program
\end{lstlisting}
Note that the ACC \code{data copy} occurs before the native Fortran ACC kernel and
after the SWIG-wrapped Thrust kernel, demonstrating that the interoperability of
the Fortran and C++ data requires no data movement.

SWIG also supports automatic conversion of MPI communicator handles between the
Fortran \code{mpi} module and MPI's C API by generating wrapper code that calls
\code{MPI\_Comm\_f2c}. The SWIG interface code
\begin{lstlisting}[language=SWIG]
%include <mpi.i>
void set_my_comm(MPI_Comm comm);
\end{lstlisting}
will generate a wrapper function that can be called from Fortran using the
standard \code{mpi} module and its communicator handles:
\begin{lstlisting}[language=Fortran]
use mpi
call set_my_comm(MPI_COMM_WORLD)
\end{lstlisting}

\subsection{Direct C binding}

A special \code{\%fortranbindc} directive in SWIG-Fortran will bypass wrapper
function generation and instead build direct \code{bind(C)} public function
interfaces in the Fortran module for C-linkage functions. A similar macro,
\code{\%fortranbindc\_type}, will generate \code{bind(C)} derived types in
Fortran from C-compatible structs, with no wrapper code. Another directive
\code{\%fortranconst} will generate \code{parameter}-qualified Fortran constants
from \code{\#define} macros, and SWIG will further generate Fortran C-bound
enumerations from C \code{enum} types.

\subsection{Language features}

The SWIG Fortran module maps many other C++ capabilities to Fortran.
Table~\ref{t:features} lists features supported by SWIG and their
implementation status in the Fortran module. SWIG currently has only limited
support for C++11 features, so these are omitted from the table.

\begin{table}[htb]
  \centering
  \caption{List of C++ features supported by SWIG and their implementation
  status (\cmark/\smark/\xmark\ for full/partial/none) in SWIG-Fortran.}
  \label{t:features}
  \begin{tabular}{lc}
    \toprule 
    Feature & Status \\
    \midrule 
    Data type conversion & \cmark \\
    \quad Fundamental types & \cmark \\
    \quad C strings & \cmark \\
    \quad Pointers and references & \cmark \\
    \quad Function pointers, member function pointers & \cmark \\
    \quad POD structs & \cmark \\
    \quad Arrays & \smark \\
    \quad Shared pointers & \cmark \\
    \quad \code{std::string} & \cmark \\
    \quad \code{std::vector} & \cmark \\
    \quad \code{thrust::device\_ptr} & \cmark \\
    \quad Other standard library containers & \smark \\
    Functions & \cmark \\
    \quad Default arguments & \cmark \\
    \quad Overloading & \smark \\
    \quad Operator overloading & \xmark \\
    Templates & \cmark \\
    Classes & \cmark \\
    \quad Member data & \cmark \\
    \quad Inheritance & \smark \\
    \quad Multiple inheritance & \xmark \\
    Constants & \cmark \\
    \quad Enumerations & \cmark \\
    \quad Compile-time constants & \cmark \\
    Exceptions & \cmark \\
    \bottomrule 
  \end{tabular}
\end{table}


\section{Applications}\label{s:examples}


This section contains two examples of the capabilities outlined in the previous
section and a discussion of their performance: \textsl{(i)} wrapping a standard C++ library
sort function, and \textsl{(ii)} accessing a sparse matrix-vector multiplication computational
kernel through a generated interface with performance discussion.

\subsection{Sorting}

Sorting arrays is a common operation in problem setup and data mapping
routines, and the C++ standard library provides generic algorithms for efficient
sorting in its \code{<algorithm>} header.  In contrast, Fortran application
developers must choose a sorting algorithm and implement it by hand for each
data type, an approach with many shortcomings. Application developers
must understand that a naive sorting algorithm that works well for desktop-sized
problems may not be performant for exascale-sized problems. Manually implementing
a robust version of a sorting algorithm is also notoriously error-prone.
Finally, having to instantiate the implemented algorithm for each data type
increases development and maintenance cost.

Using an externally supplied, efficient, and well-tested algorithm is clearly a
better approach.
The following self-contained SWIG interface wraps the C++-supplied \code{sort} implementation
into a generic Fortran subroutine that operates on either integer or real Fortran arrays:
\begin{lstlisting}[language=SWIG,numbers=right,numbersep=-10pt,stepnumber=1,numberstyle=\tiny,firstnumber=1]
%module algorithm
%{
#include <algorithm>
%}
%inline %{
template<class T>
void sort(T *ptr, size_t size) {
  std::sort(ptr, ptr + size);
}
%}
%include <typemaps.i>
%apply(SWIGTYPE *DATA, size_t SIZE)
  {(int    *ptr, size_t size),
   (double *ptr, size_t size)};
%template(sort) sort<int>;
%template(sort) sort<double>;
\end{lstlisting}
The first line declares the name of the resulting Fortran module. Lines 2--4
insert the standard library include into the generated C++ wrapper code, and the
following \code{\%inline} block both \emph{inserts} the code into the wrapper
and \emph{declares} it to SWIG. Lines 11--14 inform SWIG that a special predefined
typemap (which treats two C++ pointer/size arguments as a single Fortran array
pointer argument; see~\S\ref{s:type_conversions}) should be applied to the
function signature of the declared
\code{sort} function. The final two lines direct SWIG to instantiate the sort
function for both integer and double-precision types.

Fortran application developers need not understand or even see any of the above
code; they merely link against the compiled SWIG-generated files and use the
wrapped function as follows:
\begin{lstlisting}[language=Fortran]
use algorithm
integer(c_int), dimension(:), allocatable :: x
integer(c_double), dimension(:), allocatable :: y
! ... Allocate and fill x and y ...
call sort(x)
call sort(y)
\end{lstlisting}

A developer might wonder whether using C++ instead of Fortran for a numeric
algorithm will slow their code, so Table~\ref{t:sort} compares the performance
of two Fortran codes that sort an array of $N$ random real numbers. The first
code implements a standard Fortran quicksort numerical
recipe~\cite{fortran90guide}, and the second calls the SWIG-wrapped C++
function. Both experiments were compiled using GCC 7.3.0 and run on a Intel Xeon E5-2620 v4
workstation, and the timings were averaged across 40 runs to remove variability.

\begin{table}
  \caption{Performance comparison of $N$-element array sort.}\label{t:sort}
  \centering
  \begin{tabular}{lrrrr}
    \toprule
    & \multicolumn{4}{c}{Time (s) for $N={}$} \\
    Implementation & $10^4$ &  $10^5$ & $10^6$ & $10^7$ \\
    \midrule
    Native Fortran quicksort      & 0.0011 & 0.0104 & 0.1132 &  1.3058 \\
    Wrapped C++ \code{std::sort}  & 0.0008 & 0.0074 & 0.0877 &  1.0189 \\
    \bottomrule
  \end{tabular}
\end{table}

The external C++ sort actually outperforms the native Fortran implementation,
likely due to its algorithmic implementation: the standard C++ \code{std::sort}
function uses \emph{introsort}, which has the algorithmic strengths of both
heapsort and quicksort \cite{musser1997}, but is more lengthy to describe and
implement. Yet with SWIG-Fortran, ``implementing'' the advanced algorithm is far
easier still.

\subsection{Sparse matrix multiplication}

The sparse matrix--vector multiplication (SpMV) algorithm is a
well-known computational kernel commonly used in linear algebra. Sparse
matrices, which have relatively few nonzero elements, are typically stored and
operated upon as memory-efficient formats such as compressed row storage
(CRS)~\cite{davis2006direct}. In CRS format, a matrix with $E$ nonzero
entries and $M$ rows is stored in three arrays. The first two,
\code{vals} and \code{col\_inds}, each have $E$ elements: the values and column
indices, respectively, of each nonzero matrix entry. A third length-$M$ array,
\code{row\_ptrs}, comprises the offset of the first nonzero entry in each row of
the matrix.

With CRS, the sparse matrix-vector multiplication algorithm consists of two
nested loops:
\begin{lstlisting}[language=Fortran]
! Given matrix A stored in CRS format
! and vectors x and y, compute y = Ax
do i = 1, M
  ! Loop over nonzero entries in row i
  do j = row_ptrs(i), row_ptrs(i+1)-1
    y(i) = y(i) + vals(j) * x(col_inds(j))
  end do
end do
\end{lstlisting}


For the SWIG-wrapped SpMV algorithm, the data associated with the matrix is stored in a
C++ class and accessed through a Fortran interface. To analyze the performance,
three interfaces with different access granularity are considered. The coarsest
provides access to the full matrix, i.e.~three arrays containing the sparse
structure. The intermediate-granularity interface returns a single row of
nonzero entries for each call, and the finest interface provides functions
to access individual elements in the matrix. The three granularities require
different numbers of calls to the SWIG interface:
the matrix-granularity interface calls C++ only three times total, the
row-level interface calls C++ twice per matrix row, and the finest calls it
twice per nonzero matrix entry.

This example uses a simple 2D Laplacian matrix corresponding to a discretization
of the Poisson equation on a 2D $n \times n$ Cartesian grid. Such a matrix has 5
diagonals with gaps between diagonals and a main diagonal on $-n$, $-1$, $0$,
$+1$, and $+n$.  The ``Standard'' column of Table~\ref{t:spmv}
presents the SpMV execution timings results for $n = 3000$.

\begin{table*}[t]
  \caption{SpMV time for 2D Laplacian matrix on a $3000 \times 3000$ grid.}
  \label{t:spmv}
  \centering
  \begin{tabular}{cccccc}
    \toprule
    \multirow{2}{*}{Interface}   & \multirow{2}{*}{Standard}  & \multicolumn{2}{c}{LTO} & \multicolumn{2}{c}{LTO (null pointer check)} \\
     \cmidrule(lr){3-4}\cmidrule(lr){5-6}
     && Library only & All & Library only & All \\
    \midrule
    Matrix (coarse)    & 0.095        & 0.095 & 0.076          & 0.095 &  0.077  \\
    Row (intermediate) & 0.184        & 0.162 & 0.067          & 0.158 &  0.163  \\
    Element (fine)     & 0.684        & 0.505 & 0.215          & 0.531 &  0.325  \\
    \bottomrule
  \end{tabular}
\end{table*}

As expected, the
computational work increases with the number of calls to the wrapper, resulting
in a factor of seven slowdown to go from the coarsest to the finest granularity.
Since the compiler is unable to
inline the C++ wrapper function into the Fortran application code, the optimizer
must make unnecessarily conservative approximations that hurt performance. One
workaround is to use link-time optimization (LTO)~\cite{gcc_lto}, which
compiles user code into an intermediate representation (IR) rather than assembly
code. The IR from multiple translation units, even from different languages, can be
inlined and optimized together during the link stage.

In the case of a C++ library bundled with a SWIG-generated Fortran interface,
LTO would be applied during the library's installation to the
original C++ code, the flattened C++ wrapper file, and the Fortran module file.
However, if SWIG-generated code is part of an application, the Fortran user code
could additionally be built with LTO as well. The ``LTO'' column of
Table~\ref{t:spmv} compares the performance of the SpMV test code for these two
hypothetical situations against the ``standard'' case of no LTO. Enabling LTO as
a library improves performance modestly for the finest-grained case, but as part
of an entire application toolchain it results in dramatic (3$\times$)
performance improvements for the fine-grain interfaces.

Note that the benefits of LTO depend on the complexity of the generated wrapper
code. Adding an assertion to check for null pointers reduced LTO-provided
performance by a factor of 1.5--2$\times$, shown in the last column of
Table~\ref{t:spmv}.  Thus, the wrapper interface writer, who has a degree of
control over the generated code, should consider the tradeoff between stability
and performance.


\section{Conclusion}\label{s:conclusion}

This article introduces a new, full-featured approach to generating modern
Fortran interfaces to C++ code\changed{, allowing C++ library developers to
easily expose their work's functionality to Fortran application developers,
and potentially improving the coupling between the two different languages in
the same application}.
By leveraging the SWIG automation tool, it
supports many C++ features critical to contemporary scientific software
libraries, including inheritance, templates,
and exceptions. Future work will demonstrate SWIG's utility in exposing the
Trilinos and \changed{SUNDIALS} numerical libraries to pre-exascale Fortran application
codes.

The developed software, examples, and performance code presented here are
available under open-source licenses at \url{https://github.com/swig-fortran}.


\section*{Acknowledgments}

This research was supported by the Exascale Computing Project (17-SC-20-SC),
a joint project of the U.S. Department of Energy's Office of Science and
National Nuclear Security Administration, responsible for delivering a capable
exascale ecosystem, including software, applications, and hardware technology,
to support the nation’s exascale computing imperative.

This research used resources of the Oak Ridge Leadership Computing Facility at
the Oak Ridge National Laboratory, which is supported by the Office of Science
of the U.S. Department of Energy under Contract No. DE-AC05-00OR22725.



\bibliographystyle{IEEEtran}
\bibliography{references}

\begin{IEEEbiography}{Seth Johnson}
specializes in high-performance computational radiation transport as
a research staff member at Oak Ridge National Laboratory. Although his
background is in nuclear engineering, with a B.S.~from Texas A\&M University and a
Ph.D.~from the University of Michigan, he finds himself working on computational
geometry and software architecture more than traditional engineering. Contact
him at johnsonsr@ornl.gov.
\end{IEEEbiography}

\begin{IEEEbiographynophoto}{Andrey Prokopenko}
  is a computational scientist at Oak Ridge National Laboratory. His research
  interests include multigrid algorithms, preconditioners for linear systems,
  and extreme scale computing. He has a BS and MS in applied mathematics from
  Moscow State University, and a Ph.D. in applied mathematics from the University
  of Houston. He is a member of the Society for Industrial and Applied
  Mathematics. Contact him at prokopenkoav@ornl.gov.
\end{IEEEbiographynophoto}


\begin{IEEEbiographynophoto}{Katherine Evans}
is a group leader and senior research staff member for the Computational
Earth Sciences Group at Oak Ridge National Laboratory. She received her Ph.D. from the Georgia
Institute of Technology and completed a postdoc at Los Alamos National Laboratory studying
implicit numerical methods for phase change and fluid flows. Kate's work includes high
resolution and scalable numerical methods development for atmosphere models and verification
and validation tools for global climate models, using multiple computer languages. She is a
member of the American Meteorological Society, American Geophysical Union, and the Society
for Industrial and Applied Mathematics. Contact her at evanskj@ornl.gov.
\end{IEEEbiographynophoto}

\end{document}